\documentstyle[11pt,graphicx]{article}

\title{ Relic Gravitational Waves
         with A Running Spectral Index
        and Its Constraints at High Frequencies}

\author{\small   M.L. Tong\thanks{mltong@mail.ustc.edu.cn}\ ,
                 Y. Zhang\thanks{yzh@ustc.edu.cn} \\
        \small Key Laboratory of Galactic and Cosmological Research\\
        \small Center for Astrophysics \\
       \small University of Science and Technology of China \\
       \small Hefei, Anhui, 230026,  China }
 \date{}

\begin{document}

\maketitle
\baselineskip=19truept

\def\vek{\vec{k}}
\renewcommand{\arraystretch}{1.5}
\newcommand{\be}{\begin{equation}}
\newcommand{\ee}{\end{equation}}
\newcommand{\ba}{\begin{eqnarray}}
\newcommand{\ea}{\end{eqnarray}}

%* {e-mail:yzh@ustc.edu.cn}
\sf

\begin{center}
\Large  Abstract
\end{center}

\begin{quote}

\large \sf
\baselineskip=19truept

We study the impact of a running index $\alpha_t$
on the spectrum of relic gravitational waves (RGWs)
over the whole range of frequency
$(10^{-18}\sim 10^{10})$ Hz
and reveal its
implications in RGWs
detections and in cosmology.
Analytical calculations  show that,
although the spectrum of RGWs on low frequencies
is less affected by $\alpha_t\ne 0$,
but, on high frequencies,
the spectrum is modified substantially.
Investigations are made toward
potential detections of the $\alpha_t$-modified RGWs
for several kinds of current and planned detectors.
The Advanced LIGO will likely be able to detect
RGWs with  $\alpha_t\ge 0$  for inflationary models
with the inflation index $\beta=-1.956$
and the tensor-scalar ratio $r= 0.55$.
The future LISA can detect RGWs
for a much broader range of  ($\alpha_t$, $\beta$, $r$),
and will have a better chance to break a degeneracy between them.
Constraints on $\alpha_t$ are estimated
from several detections and cosmological observations.
Among them, the most stringent one is from the bound of
the Big Bang  nucleosynthesis (BBN),
and requires $\alpha_t < 0.008$
rather conservatively  for any reasonable  ($\beta$, $r$),
preferring  a nearly power-law spectrum of RGWs.
In light of this result,
one would expect the scalar running index  $\alpha_s$
to be of the same magnitude as $\alpha_t$,
if both RGWs and scalar perturbations are
generated by the same scalar inflation.

\end{quote}


\begin{center}
{\bf 1. Introduction}
\end{center}

Inflationary models predict a stochastic background of relic
gravitational waves (RGWs)
\cite{grishchuk,grishchuk07,starobinsky,zhang2,Maggiore,Giovannini},
whose spectrum depends upon the initial condition when they were
generated. After that, only the expansion of spacetime background
will substantially affect its evolution behavior in a determined
fashion, since the interaction of RGWs with other cosmic components
is typically very weak. Therefore, RGWs carry a unique information
of the early Universe, and serve as a probe into the Universe much
earlier than the cosmic microwave background (CMB).

Unlike gravitational waves radiated by usual astrophysical process,
RGWs exist everywhere and anytime,
and, moreover,
its spectrum spreads over a broad range of frequency,
$(10^{-18}\sim 10^{10})$ Hz.
Therefore,
RGWs are one of the major scientific goals of various GW detectors,
including the ground-based interferometers,
such as the ongoing LIGO \cite{ ligo1}, Advanced LIGO \cite{ligo2},
VIRGO \cite{virgo}, GEO \cite{geo},
and the space interferometers, such as the future LISA \cite{lisa,lisa2},
DECIGO \cite{decigo}, and  ASTROD \cite{astrod},
the cryogenic resonant bar detectors,
such as EXPLORER \cite{Astone2007}, NAUTILUS  \cite{NAUTILUS},
the cavity detectors  MAGO \cite{Ballatini},
the waveguide detector \cite{cruise},
and the proposed Gaussian maser beam detector at GHz\cite{fangyu}.
Besides, a long period of observations of
pulsar arrival times can be used as GW detectors at nanoHertz
\cite{Sazhin,Kaspi}, such as PPTA \cite{PPTA}.
Furthermore,  the very low frequency portion of
RGWs also contribute to the CMB anisotropies
and polarizations  \cite{basko},
yielding a magnetic type polarization of CMB
as a distinguished signal of RGWs.
WMAP \cite{Peiris, Spergel,Spergel07,page,Komatsu,Hinshaw,Dunkley},
Planck \cite{Planck},  and the proposed CMBpol \cite{CMBpol}
are of this type.
Therefore, the detailed information of RGWs is
much desired  both for the detection of RGWs itself
and for cosmology as well.

The spectrum of RGWs depends on several physical factors.
After being generated, RGWs will be affected
by a sequence of stages of cosmic expansion,
including the current acceleration \cite{zhang2},
also by physical processes in the early Universe,
such as the neutrino free-streaming \cite{Weinberg,yuki,Miao},
the QCD transition,
and the $e^+ e^-$ annihilation \cite{Schwarz,SWang}, etc.
But, over all, it depends most sensitively upon the initial condition,
which includes the initial amplitude,
the spectral index $\beta$, and the running index $\alpha_t$, as well.
These  parameters are predicted
by specific models of inflation scenarios  \cite{zhang2}.

In this paper, based on
our previous analytic work of RGWs  \cite{zhang2,Miao,SWang},
we will study the modifications by a non-zero $\alpha_t$,
together with $\beta$, upon the spectrum,
and explore its implications for gravitational wave detections.
By specifying the scale factor for various expanding stages
and by a normalization
in light of CMB temperature anisotropies of WMAP 5-year
\cite{Komatsu,Hinshaw,Dunkley},
the analytic spectrum $h(k, \tau_H)$ of RGWs with
dependence on $\alpha_t$ and $\beta$ will be  demonstrated.
Although a small $\alpha_t$ seems to have insignificant
influences upon the spectrum of RGWs on very low frequencies,
it will cause increasingly substantial modifications upon the spectrum
on higher frequencies.
This inevitably leads to far-reaching consequences to RGWs detection,
since most of detectors are, or will be,
operating at various medium and high frequencies,
from around  $10^{-9}$ Hz up to around  $10^{9}$ Hz.
The previously estimated constraints on RGWs
should be revised in the presence of $\alpha_t$ accordingly.
To this end, comparisons will be carried out
between the theoretical spectrum of RGWs and
the sensitivity of various ongoing and forthcoming GW detectors.
Thereby,  constraints on $\alpha_t$ will be derived
and their implications in cosmology will be discussed.

The outline of this paper is as follows.
In section 2,
the scale factor $a(\tau)$ is specified
for consecutive stages of cosmic expansion,
and the construction is briefly reviewed
for the analytical solution of the RGWs.
In section 3, we present the resulting spectrum of RGWs
with a scalar running index $\alpha_t$
and demonstrate the induced  modifications.
In section 4,
comparisons are made between the calculated RGWs and
the sensitivity of several kinds of ongoing and planned GW detectors,
thereby, constraints on RGWs are obtained
and implications in cosmology are discussed.
In this paper we use unit with $c=\hbar=k_B=1$.

\begin{center}
{\bf 2. Analytical Solution of RGWs in Expanding Universe }
\end{center}

For a spatially flat ($k=0$)  universe
the Robertson-Walker spacetime has a metric
\be
ds^2=a^2(\tau)[-d\tau^2+\delta_{ij}dx^idx^j],
\ee
where $\tau$ is the conformal time, and the scale factor  $a(\tau)$
is determined by the Friedmann equation
\be \label{Friedmann}
(\frac{a'}{a^2})^2=\frac{8\pi G}{3} \rho,
\ee
where $'  \equiv d/d\tau$.
From the very early inflation up to the present accelerating expansion,
$a(\tau)$ can be described by the following successive stages
\cite{grishchuk07,zhang2}:

The inflationary stage:
\be \label{inflation}
a(\tau)=l_0|\tau|^{1+\beta},\,\,\,\,-\infty<\tau\leq \tau_1,
\ee
where the inflation index $\beta$ is an important model parameter,
related to the spectral index $n_s$ of primordial
perturbation via
$n_s=2\beta+5$.
The special case of $\beta=-2$ corresponds the exact de Sitter expansion.
But both the model-predicted and the observed results, such as WMAP,
indicate that the value of $\beta$ can differ slightly from  $-2$.
In our presentation,
$\beta= -2.015$ and $\beta=-1.956$,
corresponding to $n_s=0.97$ and $n_s=1.089$ respectively,
are also taken for illustration.

The reheating stage \cite{Miao,SWang}:
\be
a(\tau)=a_z|\tau-\tau_p|^{1+\beta_s},\,\,\,\,\tau_1\leq \tau\leq \tau_s.
\ee
As a reheating model parameter, we will mostly take $\beta_s= -0.3$.

The radiation-dominant stage :
\be \label{r}
a(\tau)=a_e(\tau-\tau_e),\,\,\,\,\tau_s\leq \tau\leq \tau_2.
\ee

The matter-dominant stage:
\be \label{m}
a(\tau)=a_m(\tau-\tau_m)^2,\,\,\,\,\tau_2 \leq \tau\leq \tau_E.
\ee

The accelerating stage up to the present time $\tau_H$  \cite{zhang2}:
\be \label{accel}
a(\tau)=l_H|\tau-\tau_a|^{-\gamma},\,\,\,\,\tau_E \leq \tau\leq \tau_H,
\ee
where $\gamma$ is a $\Omega_\Lambda$-dependent  parameter.
For instance, $\gamma\simeq 1.06$ for $\Omega_{\Lambda}=0.65 $,
and $\gamma\simeq 1.044$ for $\Omega_{\Lambda}=0.75 $ \cite{Miao}.
To be specific, we take $\gamma\simeq 1.044$ and $\Omega_{\Lambda}=0.75 $
in this paper.

In Eqs. (\ref{inflation}) -- (\ref{accel}),
the five instances of time,
$\tau_1$, $\tau_s$, $\tau_2$, $\tau_E$, and $\tau_H$,
separate the different stages,
and can be determined by the relations \cite{Miao}:
$\frac{a(\tau_s)}{a(\tau_1)}=300$
for the reheating stage,
$\frac{a(\tau_2)}{a(\tau_s)}=10^{24}$ for the radiation stage,
$\frac{a(\tau_E)}{a(\tau_2)}
             =\frac{a(\tau_H)}{a(\tau_2)} \frac{a(\tau_E)}{a(\tau_H)}
=3454  \frac{a(\tau_E)}{a(\tau_H)}  $ for the matter stage,
and
$\frac{a(\tau_H)}{a(\tau_E)} =(\frac{\Omega_\Lambda}{\Omega_m})^{1/3}$ for
the present accelerating stage,
and $\tau_H$ is to be fixed by the normalization
\be \label{norm}
|\tau_H-\tau_a|=1.
\ee
In the expressions of $a(\tau)$,
there are twelve parameters,
among which  $\beta$, $\beta_s$ and $\gamma$
are imposed as the model parameters.
By the continuity of $a(\tau)$ and of $a(\tau)'$
at the four instances $\tau_1$, $\tau_s$, $\tau_2$ and $\tau_E$,
one can fix other eight parameters.
The remaining $l_H$  can be fixed by
\be
l_H=\gamma/H_0,
\ee
where $H_0$ is the present Hubble constant.
We will take the Hubble parameter $h_0\simeq 0.71$.
Thus  $a(\tau)$ is completely fixed \cite{Miao}.

In the present universe
the physical frequency for a conformal wavenumber  $k$ is given by
\be \label{12}
\nu=  \frac{k}{2\pi a (\tau_H)} = \frac{k}{2\pi l_H}.
\ee
 The comoving wavenumber $k_H$  corresponding to
a wavelength of Hubble radius $1/H_0$ at present is given by
\be
k_H = \frac{2\pi a(\tau_H)}{1/H_0}=2\pi \gamma,
\ee
and another wavenumber which will be used is
\be\label{ke}
k_E \equiv \frac{2\pi a(\tau_E)}{1/H(\tau_E)}
    ={k_H} (\Omega_m / \Omega_\Lambda)^{1/3\gamma},
\ee
whose corresponding wavelength at the time $\tau_E$
is equal to the Hubble radius $1/H(\tau_E)$ at that moment.
 Note that, in Eq. (\ref{ke})
 we have made corrections to that in Ref. \cite{Miao}.

In the presence of
the gravitational waves,
the perturbed metric  is
\be
ds^2=a^2(\tau)[-d\tau^2+(\delta_{ij}+h_{ij})dx^idx^j],
\ee
where the tensorial perturbation $h_{ij}$
is traceless $h^i_{\,\,i}=0$ and transverse $h_{ij,j}=0$.
It can be decomposed into the Fourier $k$-modes and into the polarization
states, denoted by $\sigma$,  as
\be
\label{planwave}
h_{ij}(\tau,{\bf x})=
   \sum_{\sigma=\times,+}\int\frac{d^3k}{(2\pi)^3}
         \epsilon^{\sigma}_{ij}h_k^{(\sigma)}(\tau)e^{i\bf{k}\cdot{x}},
\ee
where
$h_{-k}^{(\sigma)*}(\tau)=h_k^{(\sigma)}(\tau)$
ensuring that $h_{ij}$ be real,
$\epsilon^{\sigma}_{ij}$ is the polarization tensor.
In terms of the mode $h^{(\sigma)}_{k}$,
the wave equation is
\be \label{eq}
h^{ (\sigma) }_{k}{''}(\tau)
+2\frac{a'(\tau)}{a(\tau)}h^{ (\sigma) }_k {'}(\tau)
+k^2 h^{(\sigma)}_k(\tau )=0.
\ee
Assuming each polarization, $\times$,  $+$,
$h^{(\sigma)}_k(\tau )$ has the same statistical properties,
the super index $(\sigma)$ can be dropped
As listed in Eq.(\ref{inflation}) through Eq.(\ref{accel}),
the scale factor has a power-law form
\be
a(\tau) \propto \tau^\alpha,
\ee
and the solution to Eq.(\ref{eq}) is a linear combination of
Bessel and Neumann functions
\be \label{hom}
h_k(\tau)=\tau^{\frac{1}{2}-\alpha}
 \big[C_1 J_{\alpha-\frac{1}{2}}(k \tau)
      +C_2   N_{\alpha-\frac{1}{2}}(k \tau)\big],
\ee
where the constants $C_1$ and $C_2$ for each stage are determined
by the continuity of $h_k$ and of $h'_k$
at the joining points
$\tau_1,\tau_s,\tau_2$ and $\tau_E$ \cite{zhang2,Miao,SWang}.
Therefore, the analytical solution  of RGWs is completely fixed,
once the initial condition during the inflation is given.

For the inflationary stage, one has
\be \label{infl}
h_k(\tau) = A_0 l_0^{-1}|\tau|^{-(\frac{1}{2}+\beta)}
\big[ A_1 J_{\frac{1}{2}+\beta}(k\tau)
     +A_2 J_{-(\frac{1}{2}+\beta)}( k\tau) \big],
\,\,\,\,\, \, -\infty<\tau\leq \tau_1,
\ee
where the $k$-independent constant $A_0$  determines the initial amplitude,
and
\be
A_1=-\frac{i}{\cos \beta\pi}\sqrt{\frac{\pi}{2}}e^{i\pi\beta/2},
\,\,\,\,\,
A_2=iA_1e^{-i\pi\beta},
\ee
are taken \cite{grishchuk1993,zhang2,Miao},
so that in the high frequency limit
$\lim_{k\rightarrow \infty}h_k(\tau)\propto e^{-ik\tau} $
the  \textit{adiabatic vacuum} is achieved \cite{parker}.
In the long wave-length limit,
$k\tau\ll 1$, the $k$-dependence of $h_k(\tau)$ is given by
\be
h_k(\tau)\propto J_{\frac{1}{2}+\beta}(k\tau )\propto
 k^{\frac{1}{2}+\beta}.
\ee

\begin{center}
 {\bf 3. Spectrum of RGWs with a Running Index}
\end{center}

The spectrum of RGWs $h(k,\tau)$ at a time $\tau$
is defined by the following equation:
\be
\int_0^{\infty}h^2(k,\tau)\frac{dk}{k}\equiv\langle0|
h^{ij}(\textbf{x},\tau)h_{ij}(\textbf{x},\tau)|0\rangle,
\ee
where the right-hand side is the expectation value of the
$h^{ij}h_{ij}$.
Calculation  yields the spectrum  as follows
\be \label{relation0}
h(k,\tau)=\frac{\sqrt{2}}{\pi}k^{3/2} |h_k(\tau)|.
\ee
Note that this expression has a factor $\sqrt 2$
in place the factor $2$ in Ref. \cite{Miao}.
As the initial  condition,
the primordial spectrum of RGWs at the time $\tau_i$
of the horizon-crossing
during the inflation is usually taken to
be a power-law form \cite{grishchuk07,zhang2,Miao}:
\be \label{initialspectrum}
h(k,\tau_i) =A(\frac{k}{k_H})^{2+\beta},
\ee
where the index $\beta \simeq -2$ for a nearly scale-invariant spectrum,
and $A$ is proportional to  $A_0$ in Eq.(\ref{infl}).
In principle,
both $\beta$ and $A$ are determined by the specific inflationary model.
Here we take them as two independent parameters.
In literature, the following notation is often used for
the RGWs spectrum \cite{Peiris,Komatsu}
\be\label{deltah}
\Delta^2_h(k)\equiv h^2(k,\tau_i)
             =\Delta^2_h(k_0)\left(\frac{k}{k_0}\right)^{n_t},
\ee
where $k_0$ is a conformal pivot wavenumber,
whose corresponding physical wavenumber
is $k^p_0=k_0/a(\tau_H)$.
For WMAP,
the pivot $k_0^p=0.002$ Mpc$^{-1}$ is
taken \cite{Peiris,Spergel,Komatsu}.
Comparing Eqs. (\ref{initialspectrum}) and (\ref{deltah}) yields
\be
n_t=2\beta+4
\ee
and
\be\label{relation}
A=\Delta_h(k_0)\left(\frac{k_H}{k_0}\right)^{  2+\beta}.
\ee
For each cosmological model with $k_H$ and $\beta$
  being given,
  $\Delta_h(k_0)$ determines $A$.
  More often in literature, a tensor-to-scalar ratio $r$ is
  introduced \cite{Peiris}
\be\label{ratio}
r\equiv\frac{\Delta^2_h(k_0)}{\Delta^2_\Re(k_0)},
\ee
where $\Delta^2_\Re(k_0)$ is the amplitude
of the curvature spectrum at $k=k_0$,
and has been fixed
$\Delta^2_\Re(k_0)=(2.41\pm 0.11)\times10^{-9}$
by WMAP5 Mean \cite{Dunkley},
and $\Delta^2_\Re(k_0)=(2.445\pm0.096)\times10^{-9}$
by WMAP5+BAO+SN Mean \cite{Komatsu}.
Note that
    this scalar amplitude has been ``fixed'' only after the assumption
   that the data do not contain gravitational waves, i.e. $r=0$.
   Here we use $r$  as a convenient representation
   of the amplitude normalization of $\Delta_h(k_0)$ at $k_0$, i.e.,
   $\Delta_h(k_0)= 4.94\times 10^{-5} r^{1/2}$.
Thus, Eq.(\ref{relation}) is rewritten as the following
\be \label{A}
A= 4.94\times 10^{-5} r^{\frac{1}{2}}(\frac{k_H}{k_0})^{2+ \beta }.
\ee
At present,
only observational constraints on $r$
have been given by WMAP \cite{Komatsu,Hinshaw,Dunkley}.
To be specific in our presentation,
 $r=0.55$ and $r=0.22$ will be taken, respectively.

In general,
the spectra of primordial perturbations, both scalar and tensorial,
deviate from the exact power-law form
 except when the inflation  potential is an exponential.
As an extension,
one usually consider the following form of
power spectra \cite{Kosowsky,Liddle}
\ba
&&\Delta^2_\Re(k)=\Delta^2_\Re(k_0)
\left(\frac{k}{k_0}\right)^{-1+n_s(k_0)
           +\frac{1}{2}\alpha_s \ln{(k/k_0)}}, \label{scalar}\\
&&\Delta^2_h(k)=\Delta^2_h(k_0)
\left(\frac{k}{k_0}\right)^{n_t(k_0)
           +\frac{1}{2}\alpha_t \ln{(k/k_0)}}, \label{tensor}
\ea
which contain the ``running'' spectral indices
$\alpha_s\equiv dn_s/d\ln k$ for the scalar perturbations
and $\alpha_t\equiv dn_t/d\ln k$ for the tensorial perturbations.
Currently, WMAP has given some preliminary constraint on
the scalar index $n_s$ and the scalar running index $\alpha_s$.
At the pivot wavenumber $k^p_0=k_0/a(\tau_H)=0.002$ Mpc$^{-1}$,
WMAP1 has given $n_s=1.20^{+0.12}_{-0.11}$
           and  $\alpha_s=-0.077^{+0.050}_{-0.052}$ \cite{Peiris};
WMAP3 has given $n_s=0.951^{+0.015}_{-0.019}$
           and $\alpha_s=-0.055^{+0.029}_{-0.035}$ \cite{Spergel07}.
With a better determination of the third acoustic peak,
WMAP5 has given an improved result:
$n_s=1.087^{+0.072}_{-0.073}$
           and $\alpha_s=-0.050 {\pm 0.034}$ \cite{Komatsu};
and WMAP5+BAO+SN has given $n_s=1.089^{+0.070}_{-0.068}$
           and $\alpha_s=-0.053^{+0.027}_{-0.028}$,
           or $n_s=0.970\pm 0.015$ without $\alpha_s$ \cite{Komatsu}.
Compared with $n_s$,
the value of the scalar running $\alpha_s$ is relatively small.
Thus WMAP5 data do not significantly
prefer a scalar running index \cite{Dunkley}.
See also Refs. \cite{Easther} for relevant discussions.

But so far there is no direct observation of
the tensorial index $n_t$ nor the running index  $\alpha_t$.
In the slow roll  inflationary models driven by a single scalar field,
the tensorial indices, $n_t$ and $\alpha_t$,
are determined by  the inflationary potential and its derivatives,
so are the scalar ones, $n_s$  and $\alpha_s$, as well
  \cite{Kosowsky,Liddle}.
There would be relations between
the tensorial indices and the scalar ones,
if one imposes further a consistency relation.
In our context, for generality,
we will treat $n_t$ and $\alpha_t$ as parameters
independent of $n_s$ and $\alpha_s$.
Corresponding to Eq.(\ref{tensor}),
the primordial spectrum in Eq. (\ref{initialspectrum})
is  modified to
\be \label{initialspectrum2}
h(k,\tau_i) = A (\frac{k}{k_H})^{2+\beta} A_{\alpha_t} (k)
\ee
where the the extra factor
\be \label{Aalphat}
A_{\alpha_t} (k) \equiv (\frac{k}{k_0})^{\frac{1}{4}\alpha_t\ln{(k/k_0)}}
\ee
is the $\alpha_t-$induced deviation from the simple power-law spectrum,
reflecting an extra bending.
With the help of Eq.(\ref{A}), this is
\be \label{initialspectrum3}
h(k,\tau_i) =  \Delta_\Re(k_0){r}^{\frac{1}{2}}
                  (\frac{k}{k_0})^{2+\beta}
                  A_{\alpha_t} (k).
\ee
For a tiny  $\alpha_t=0.01$,
in the very low frequency range,
from $\nu=\nu_0 =3\times 10^{-18}$ Hz to $\nu =3\times 10^{-16}$ Hz,
 $ A_{\alpha_t} (k)$ goes from 1 to $1.055$,
causing only a minor increase in the amplitude by $\le 5.5\%$.
However, at very high frequency, say $\nu = 10^{9}$ Hz,
$ A_{\alpha_t} (k)\sim 10^4$,
enhancing the amplitude by 4 orders of magnitude.
This drastic effect requires a detailed investigation
into $\alpha_t$ and its consequential implications.

As discussed in Refs. \cite{zhang2,Miao,SWang},
 at the present time $\tau=\tau_H$,
the long wavelength modes $h_k(\tau)$ with wavenumber $k\leq k_E$
are still outside the horizon,
their spectrum are still of the form in (\ref{initialspectrum3}),
$h(k,\tau_H) = h(k,\tau_i) $.
Given this initial condition with a running index,
the analytic calculation of the spectrum of RGWs
can be carried out straightforwardly,
in the same way  as the non running case \cite{zhang2,Miao}.
The only difference in the actual computing procedure
is the amplitude normalization of the  RGWs spectrum,
which can be taken at the wavenumber $k_E$,
corresponding to a physical frequency
$\nu_E = k_E/2\pi a(\tau_H)=H_0/(1+z_E)\sim  10^{-18}$ Hz.
With the help of Eq.(\ref{initialspectrum3}),
it is given by
\be \label{alpha}
h(k_E,\tau_H)=  \Delta_\Re(k_0){r}^{\frac{1}{2}}
    (\frac{k_E}{k_0})^{2+\beta+\frac{1}{4}\alpha_t\ln{(k_E/k_0)}}.
\ee
In a cosmological model with a given
set of  $(r, \beta, \alpha_t)$,
the resulting  RGWs spectrum $h(\nu,\tau_H)$ at present
 is fully determined.

\begin{figure}
\centerline{\includegraphics[width=10cm]{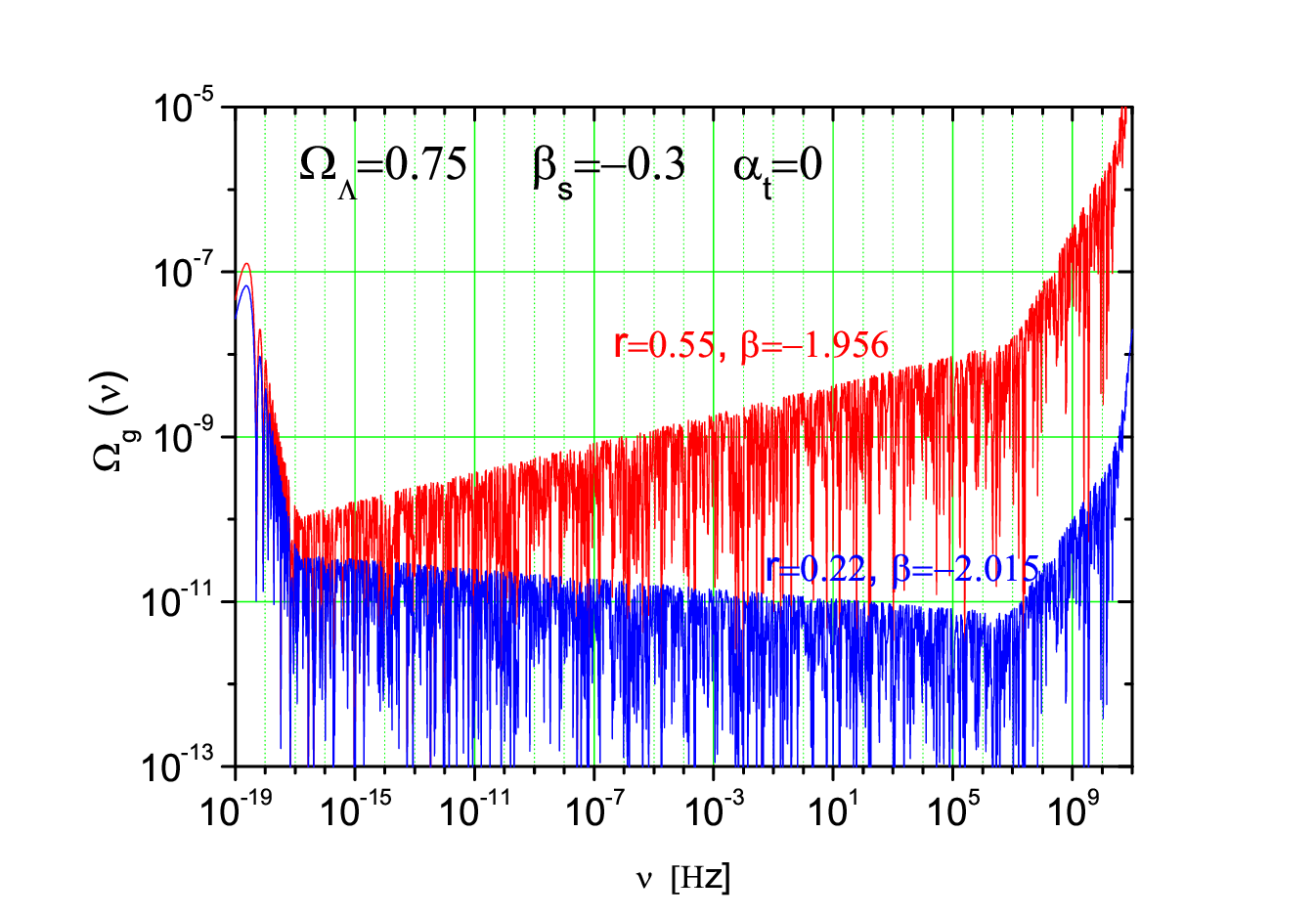}}
\caption{\label{diff_beta_Omega}
The spectral energy density
$\Omega_{g}(\nu)$ with $\alpha_t=0$ for two sets
of  $r$ and $\beta$. }
\end{figure}

Another  important quantity often used in constraining
RGWs is its present energy density parameter defined by
$ \Omega_{gw}=\frac{\rho_{g}}{\rho_c} $,
where $\rho_g=\frac{1}{32\pi G}h_{ij,0}h^{ij}_{,0}$
is the energy density of RGWs,
and $\rho_c=3H_0^2/8\pi G$ is the critical energy density.
A direct calculation yields \cite{grishchuk07}
\be\label{gwe}
\Omega_{gw}=
\int_{\nu_{low}}^{\nu_{upper}} \Omega_{g}(\nu)\frac{d\nu}{\nu},
\ee
with
\be\label{32}
\Omega_{g}(\nu)=\frac{\pi^2}{3}
        h^2(\nu,\tau_H)
     \Big(\frac{\nu}{\nu_H}\Big)^2
\ee
being the dimensionless  spectral energy density.
From this expression,  it is seen that
the spectral energy density $\Omega_{g}(\nu)$
and the spectrum $h(\nu,\tau_H)$ are two equivalent quantities,
and $h(\nu,\tau_H)/\sqrt{2}$ is just the characteristic amplitude,
denoted by $h_c(f)$   in Ref.\cite{Maggiore}.
As the cutoffs  of frequencies,
the lower and upper limit of integration in Eq.(\ref{gwe})
can be taken to be  $\nu_{low}\simeq2\times10^{-18}$ Hz
and  $\nu_{upper}\simeq10^{10}$ Hz, respectively  \cite{Miao}.
In the case of $\alpha_t=0$,
the slope of $\Omega_{g}(\nu)$ is fixed by the index $\beta$,
and the overall amplitude of $\Omega_{g}(\nu)$ is fixed by $r$,
as shown in Fig. \ref{diff_beta_Omega}.
In the following,   two combinations
$(r=0.55, \beta=-1.956)$ and $(r=0.22, \beta=-2.015)$
will often be taken for specific illustrations.

\begin{figure}
\centerline{\includegraphics[width=10cm]{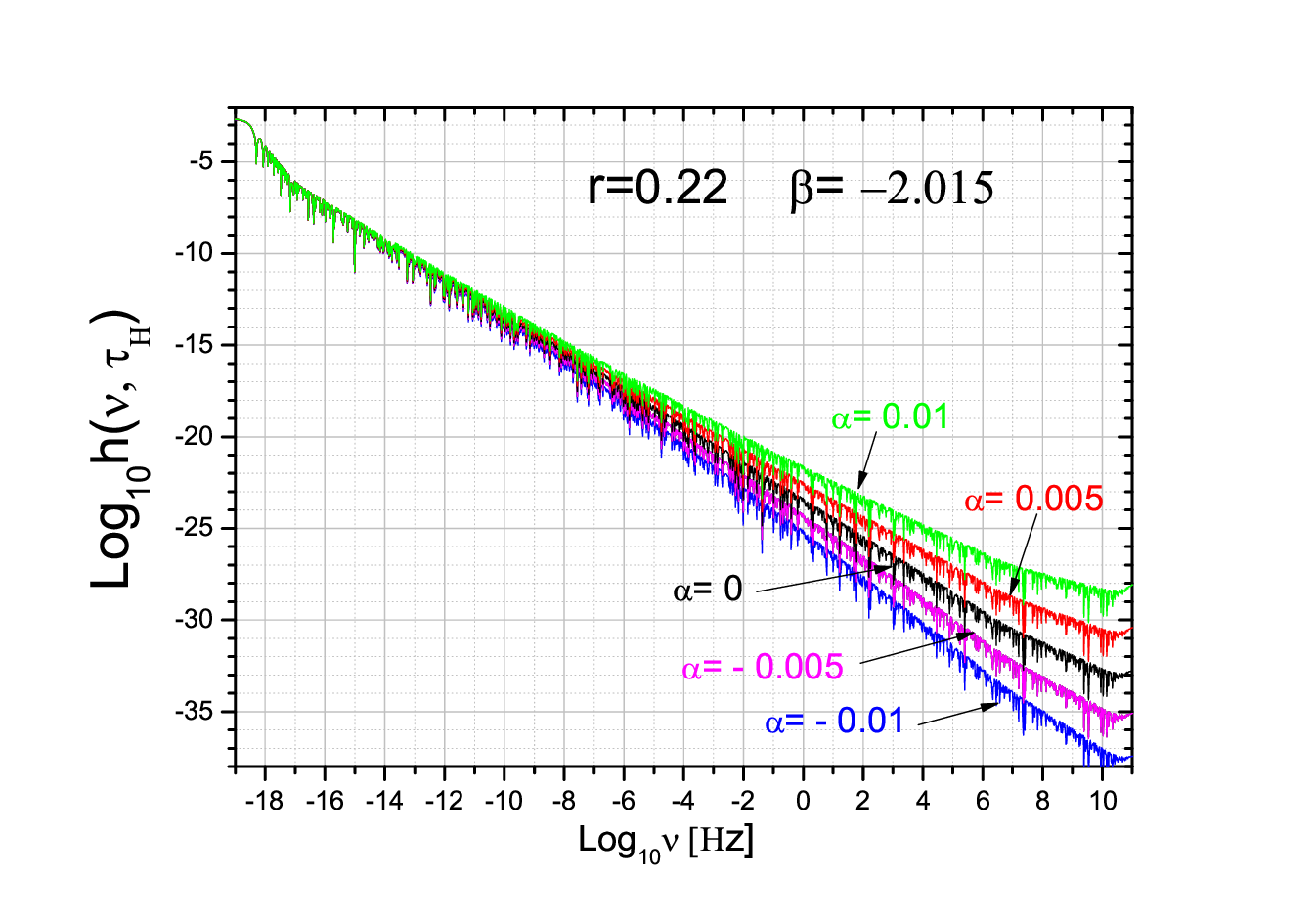}}
\caption{\label{sepc22}
The spectrum of RGWs  for various  $\alpha_t$.}
\end{figure}

\begin{figure}
\centerline{\includegraphics[width=10cm]{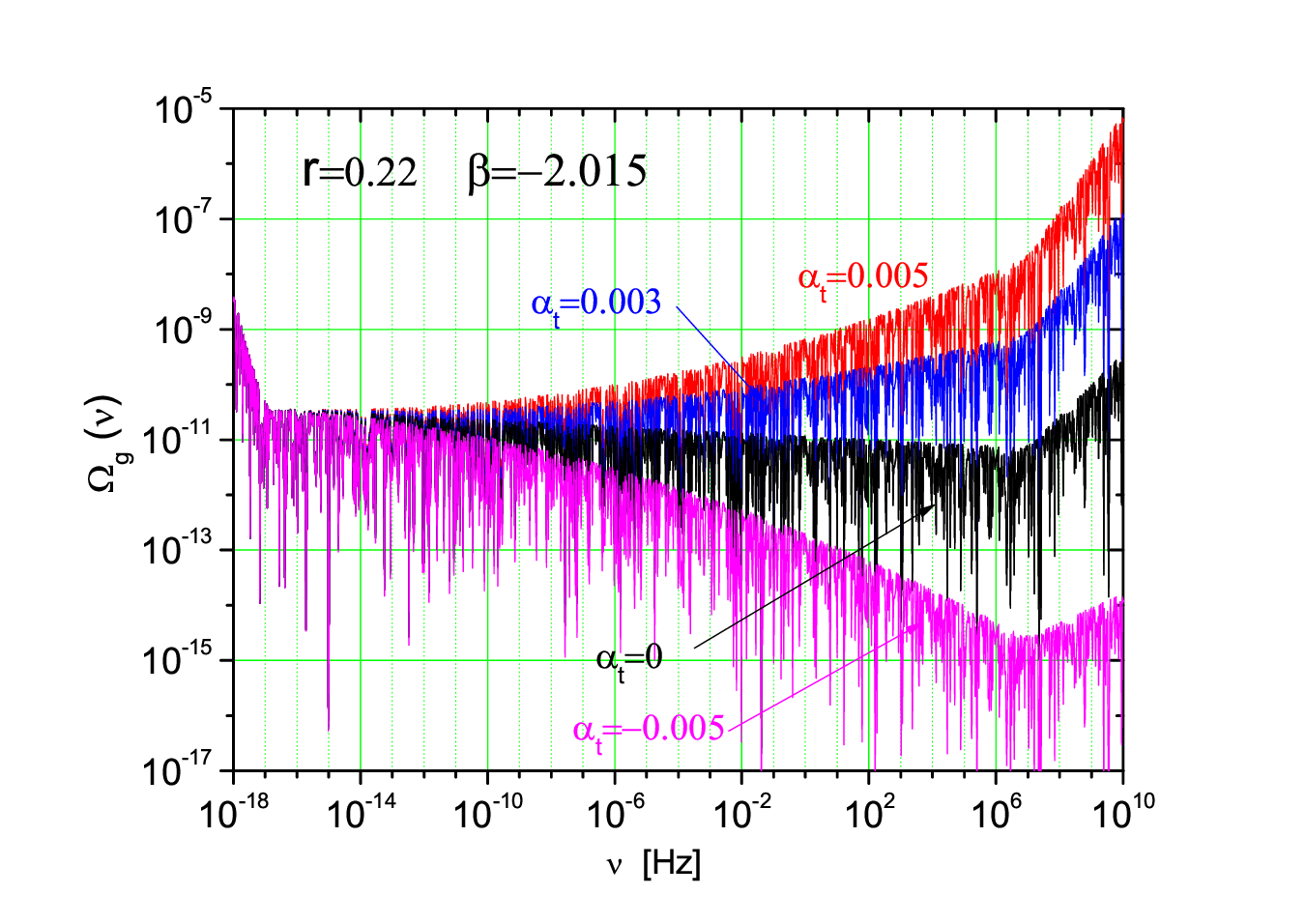}}
\caption{\label{Omegag}
The spectral energy density
$\Omega_{g}(\nu)$ for various  $\alpha_t$. }
\end{figure}

When there is a running index $\alpha_t\ne 0$,
the slopes of $\Omega_{g}(\nu)$,  and of $ h(\nu,\tau_H)$ as well,
are affected substantially.
Fig. \ref{sepc22} demonstrates  $h(\nu,\tau_H)$
for various values of $\alpha_t$ in the model with
$r=0.22$ and $\beta = -2.015$.
It is seen that
a greater $\alpha_t$ yields a higher $h(\nu,\tau_H)$
and that the modifications due to $\alpha_t$
increase with the frequency $\nu$.
While the effects are small in the  low frequency range,
they are quite substantial in the high-frequency range.
For instance,  in going from $\alpha_t=-0.01$ to $\alpha_t= 0.01$,
the amplitude of $h(\nu,\tau_H)$ gets
enhanced by $3$ orders of magnitudes
at $\nu\simeq 10^{-2}$ Hz falling the range for LISA,
$5$ orders at $\nu\simeq 10^{2}$ Hz for LIGO,
$6$ orders at $\nu\simeq 10^4$ Hz for MAGO \cite{Ballatini},
and $9$ orders at $\nu\simeq 10^{9}$Hz
for the Gauss beam \cite{fangyu}.
Equivalently,
Fig.\ref{Omegag} gives the  $\alpha_t-$dependence of
the spectral energy density $\Omega_{g}(\nu)$,
which shows more drastically
the variations in high frequencies due to $\alpha_t$.

Notice that the ratio $r$, the index $\beta$,
and the running index $\alpha_t$
in Eq.(\ref{initialspectrum2}) play
different roles in shaping the spectrum of RGWs.
$r$ sets the amplitude,
$\beta$ fixes the overall slope of the spectrum,
and $\alpha_t$ gives an extra  bending to it.
However,  in a rather narrow interval of detecting frequencies,
$r$, $\beta$,  and $\alpha_t$ have a degeneracy to certain extent,
since a larger value of each of them
tends to enhance the amplitude of $h(\nu,\tau_H)$ in the interval.
The narrower the interval is,
the stronger the degeneracy will be.
Therefore, if  a detector is operating
in a narrow interval of frequencies,
it can only detect RGWs for a combination of
parameters $(r, \beta, \alpha_t)$.
Those detectors operating over a broad frequency interval
will have a better chance to break the degeneracy.

\begin{center}
 {\bf 4. Constraints from Detections and Implications }
\end{center}

The $\alpha_t$-induced modification of the  RGWs spectrum
has practical implications for
the ongoing and planned GW detections
in the medium, and the high-frequency ranges.
Some  previously estimated constraints on RGWs
were based on the theoretical spectrum without a running index
\cite{zhang2,Miao,SWang}.
Now these will be subsequently revised to certain extent.

\begin{figure}
\centerline{\includegraphics[width=10cm]{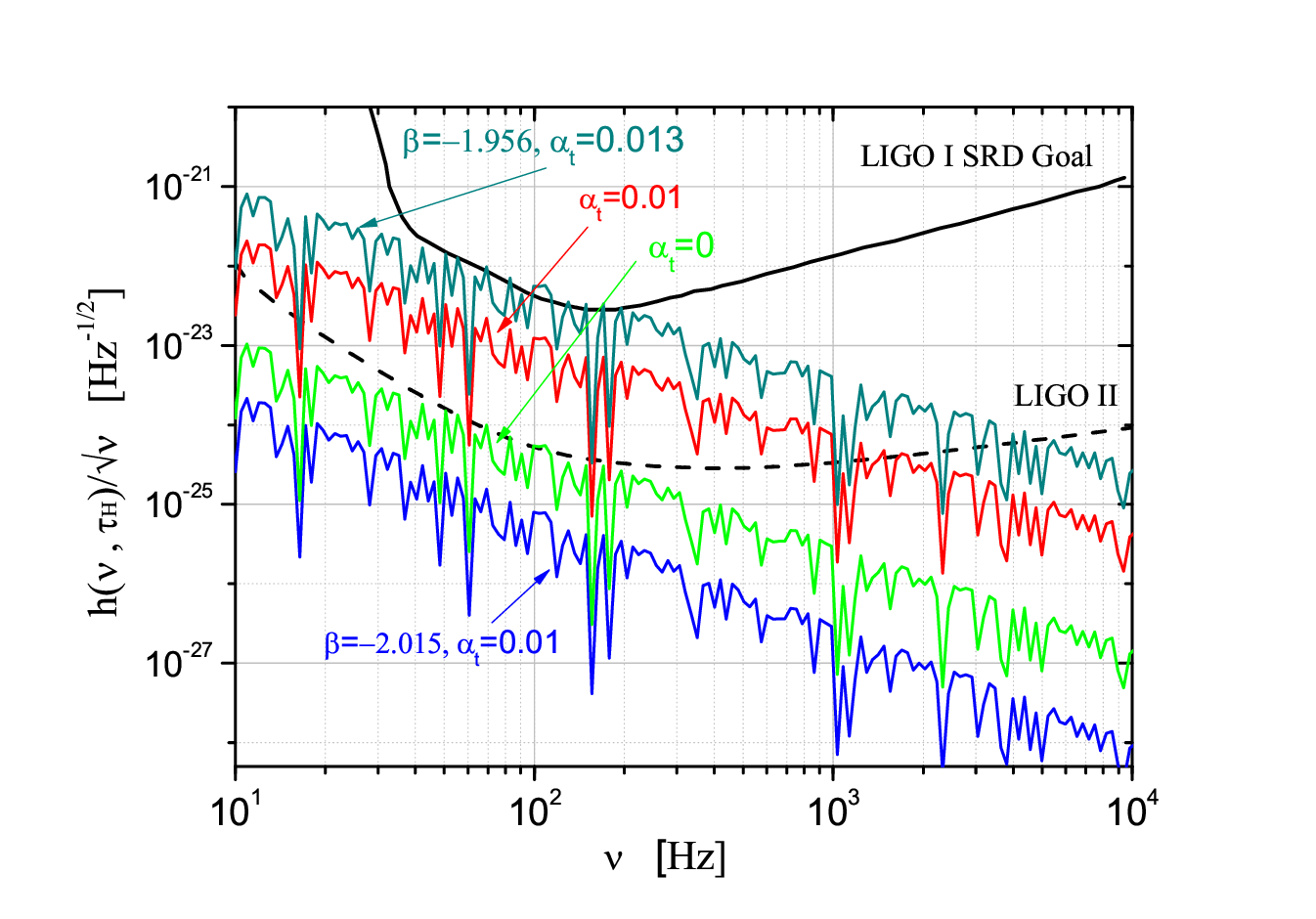}}
\caption{\label{ligo_sensitivity}
 Comparisons of the spectra with the sensitivity
of the LIGO I SRD Goal achieved by LIGO S5  ~\cite{ligo1}
and of the Advanced LIGO \cite{ligo2}.
}
\end{figure}

Figure~\ref{ligo_sensitivity} gives the comparison
of  the sensitivity curves of LIGO \cite{ligo1} and Advanced LIGO \cite{ligo2}
with  the theoretical spectra of RGWs
for various parameters  ($r$,  $\beta$, $\alpha_t$),
in the frequency range $(10^{1}, 10^4)$ Hz.
Note that, in order to compare to
the strain sensitivity $\tilde{h}_f(\nu)$ \cite{Maggiore,Allen}
of the  detectors,
the amplitude per root Hz, ${h(\nu,\tau_H)}/{\sqrt{\nu}}$,
has been used.
It is seen that,
for $r=0.55$ and $\beta=-1.956$,
the LIGO I SRD  \cite{ligo1} has already put a constraint
on the running index: $\alpha_t\le 0.013$.
It will yet not be able to detect the signals of RGWs
for $r=0.55$,  $\beta < -1.956$ and $\alpha_t <0.013$.
On the other hand,
 with a substantial improvement in sensitivity,
Advanced LIGO
will be able to detect the RGWs from models
with $r=0.55$ and $\beta> -1.956$ and $\alpha_t>0$,
but still it will unlikely be able to detect
RGWs for  $r=0.22$ and $\beta=-2.015$ even for $\alpha_t=0.01$
and less.

\begin{figure}
\centerline{\includegraphics[width=10cm]{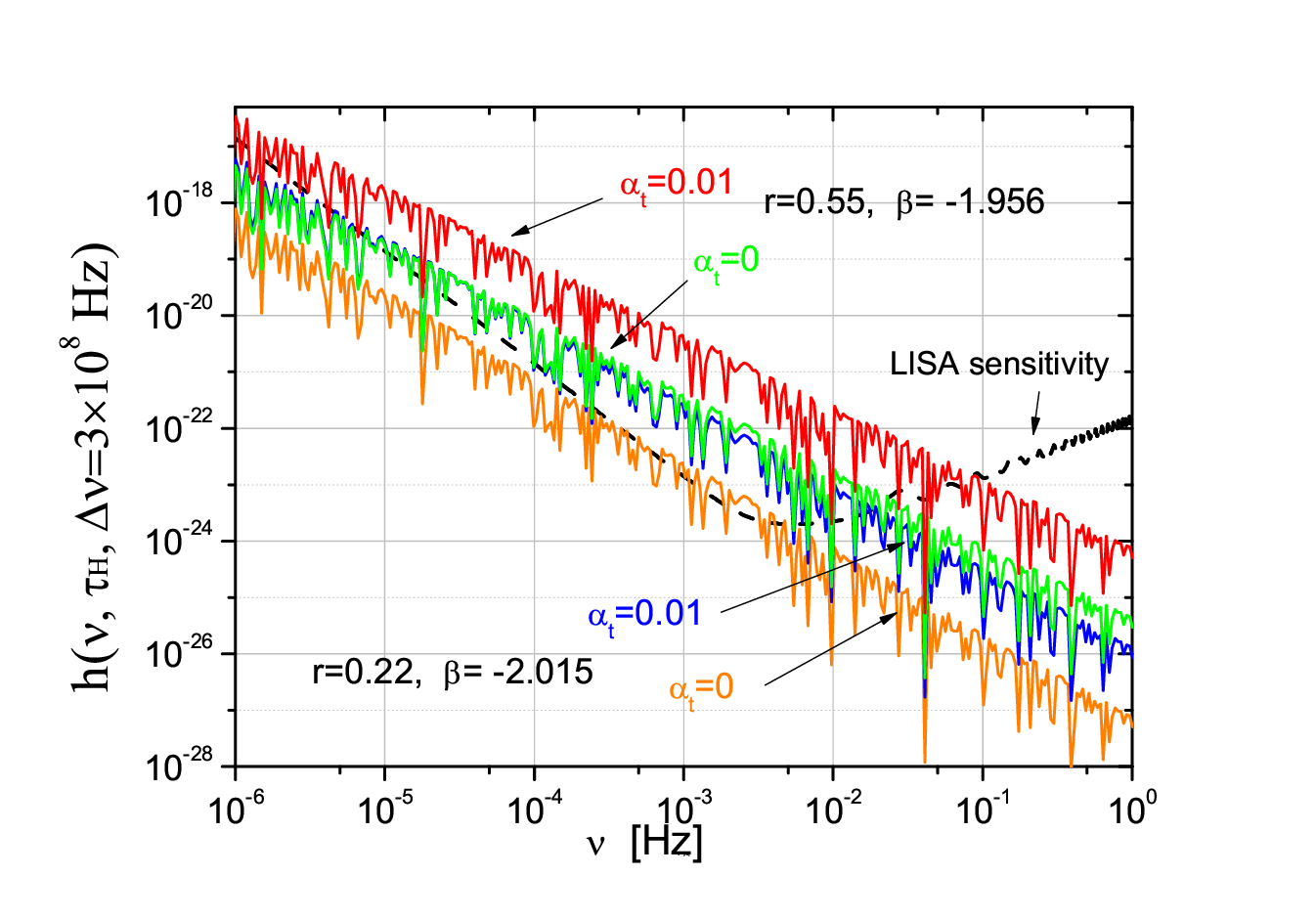}}
\caption{\label{lisa_sensitivity}
Comparison of the rms spectrum with
the LISA sensitivity curve~\cite{lisa2}.}
\end{figure}

\begin{figure}
\centerline{\includegraphics[width=10cm]{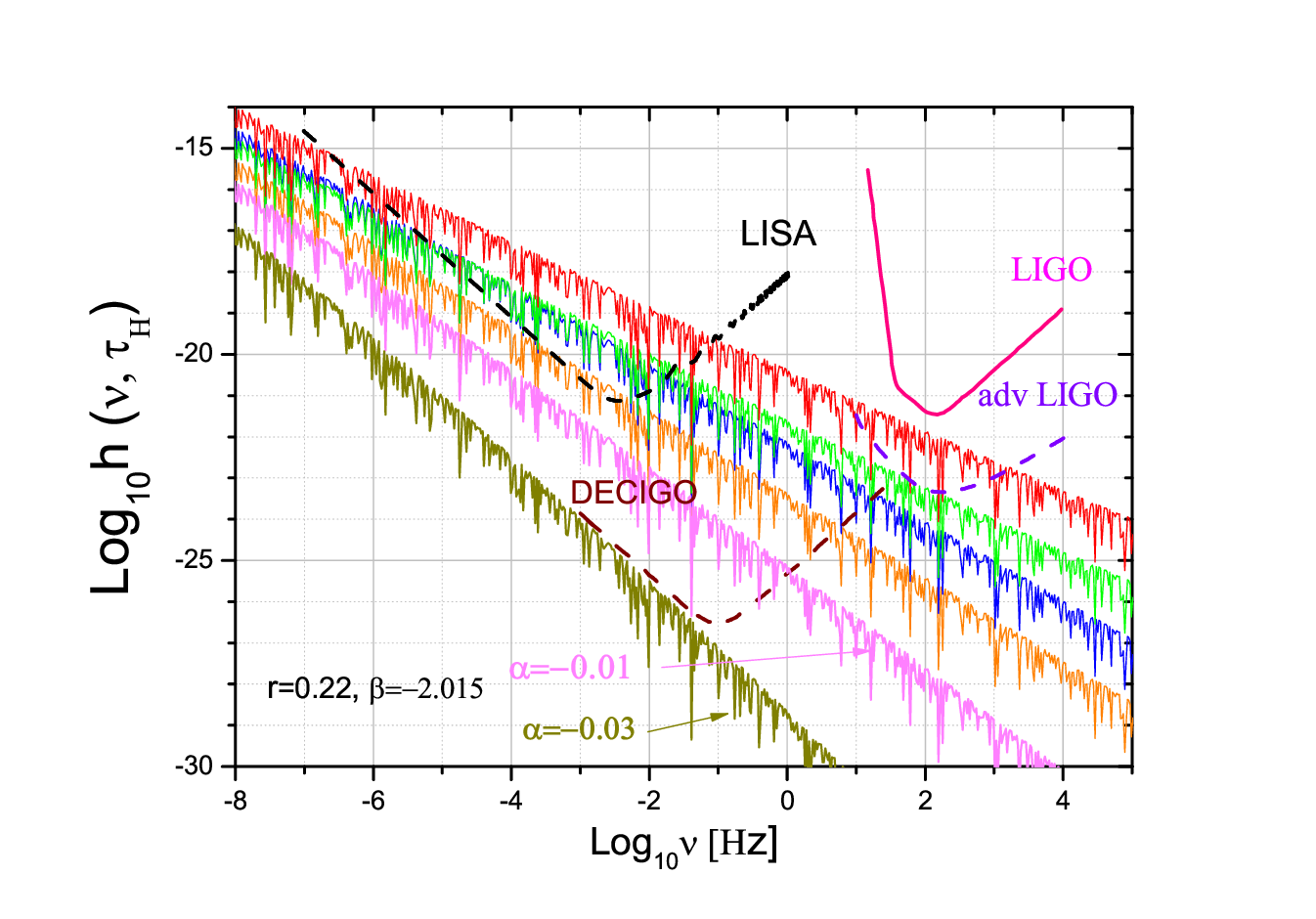}}
\caption{\label{LIGO LISA}
 Comparison of the theoretical spectrum $h(\nu,\tau_H)$ with
the converted sensitivity curve of LIGO, Advanced LIGO,  LISA,
and DECIGO \cite{decigo}.
The parameters of the four top spectra $h(\nu,\tau_H)$
are the same as the previous Fig.\ref{lisa_sensitivity}.  }
\end{figure}

Figure~\ref{lisa_sensitivity}
is a comparison of the theoretical spectrum
with the LISA  sensitivity curve \cite{lisa,lisa2}
for the ratio $S/N=1$ in the frequency range $(10^{-7}, 10^0)$ Hz,
where one year observation time has been assumed,
which corresponds to a frequency bin
$\Delta\nu \simeq 3\times 10^{-8}$Hz
around each frequency.
To make a comparison with the sensitity curve,
one needs to rescale  the theoretical spectrum $h(\nu,\tau_H)$
into the rms spectrum $h(\nu,\tau_H,\Delta\nu)$
in the band $\Delta \nu$ \cite{grishchuk07,Maggiore},
\be \label{rmssp}
h(\nu,\tau_H, \Delta\nu)
= h(\nu,\tau_H)\sqrt{\frac{\Delta\nu}{\nu}}.
\ee
The plot shows that
LISA by its present design will be quite effective in detecting
the RGWs around a range of frequencies ($10^{-6},1.5\times 10^{-1}$) Hz.
In particular, the plot tells that LISA will be able to
detect RGWs for parameters $r \ge 0.22$, $\beta \ge -2.015$,
and $\alpha_t \ge 0$.
Thus, regarding to detection of RGWs,
LISA is expected to perform much better than LIGO and  Advance LIGO.
This advantage is due to the property that
the RGWs have a higher amplitude in the range of lower frequencies.
The situation is illustrated in Fig.{\ref{LIGO LISA}},
in which the sensitivity curves of LIGO and LISA
are converted pertinently,
in order to compare with the theoretical spectrum $h(\nu,\tau_H)$.
Moreover, regarding to the RGWs detection,
the frequency range covered by LISA is very broad,
as compared with LIGO.
This is because the low frequency portion of  LISA sensitivity curve
has a slope that is rather close to that of RGWs spectrum
in the involved region.
This feature of LISA is important and can be instrumental
in breaking the $(r, \beta, \alpha_t)$ degeneracy,
as mentioned earlier.
Besides, in Fig \ref{LIGO LISA}
the sensitivity of planned DECIGO \cite{decigo}
is also presented,
together with two more spectra calculated
for very low running indices $\alpha_t=-0.01$ and $\alpha_t=-0.03$,
in the model $(r=0.22,\beta=-2.015)$, respectively.
DECIGO, if implemented, will be a powerful detector,
capable of detecting RGWs with
a very low running index $\alpha_t>-0.03$.


\begin{figure}
\centerline{\includegraphics[width=10cm]{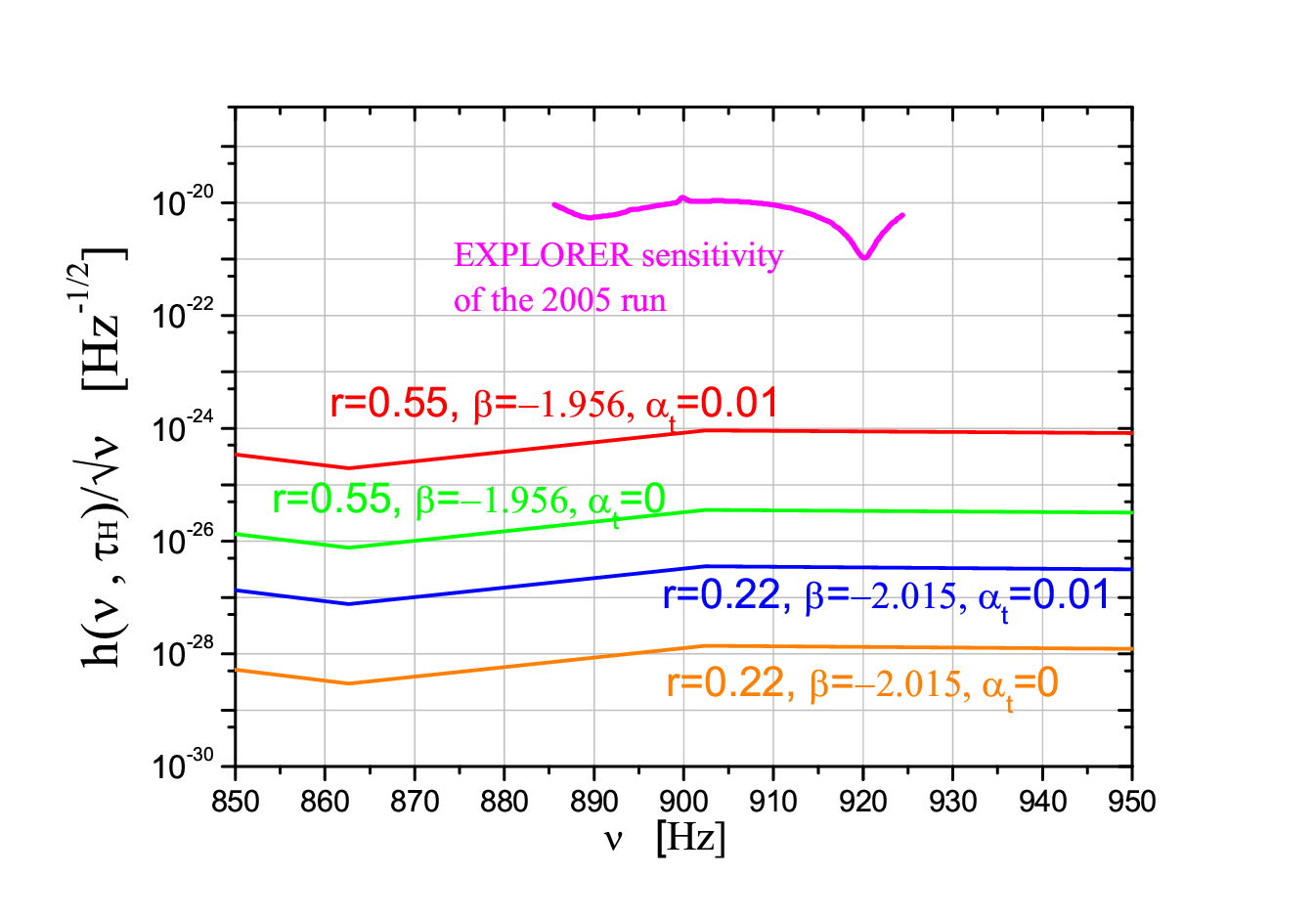}}
\caption{\label{Explorer}
Comparisons of the spectrum with
the EXPLORER sensitivity curve \cite{Astone2007}.}
\end{figure}

Fig.{\ref{Explorer}} compares
the theoretical spectrum ${h(\nu,\tau_H)}/{\sqrt{\nu}}$
with the 2005 run sensitivity curve of cryogenic resonant bar detector,
EXPLORER
in the frequency range $(890, 920)$ Hz \cite{Astone2007}.
It is seen that, even for $\beta=-1.956$ and $\alpha_t=-0.01$,
RGWs are still far beyond the reach of EXPLORER by three orders of magnitude.

\begin{figure}
\centerline{\includegraphics[width=10cm]{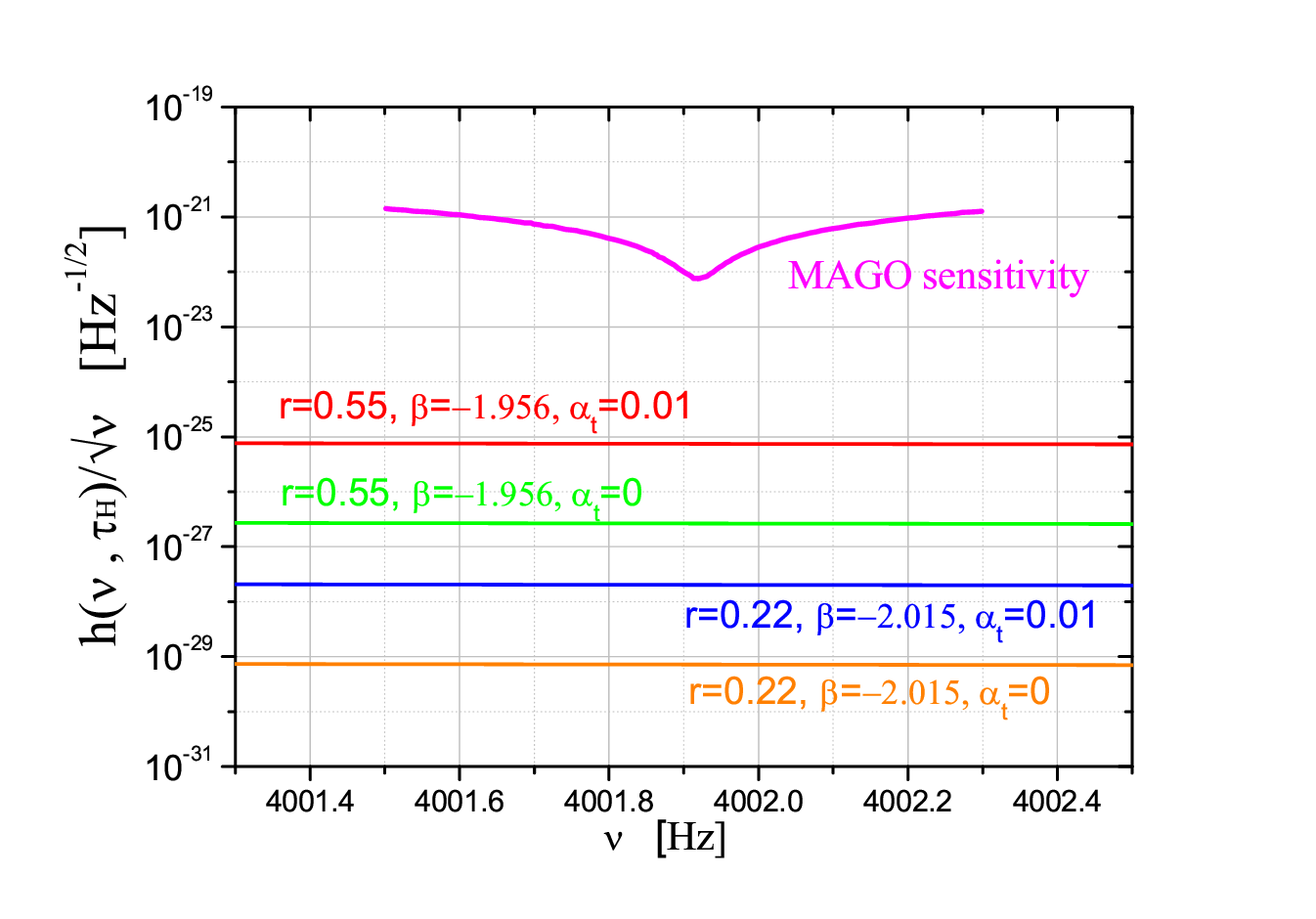}}
\caption{\label{MAGO}
Comparisons of the spectrum
with the MAGO sensitivity curve \cite{Ballatini}.}
\end{figure}

Fig.{\ref{MAGO}} compares
the theoretical spectrum ${h(\nu,\tau_H)}/{\sqrt{\nu}}$
with the sensitivity curve of MAGO,
a double spherical cavity detector,
around the frequency  $4002$ Hz  \cite{Ballatini}.
For $\beta=-1.956$ and $\alpha_t=0.01$,
RGWs are also beyond the reach of MAGO by three orders of magnitude.

The proposed Gaussian maser beam detector will operate
at a very high frequency, say $\sim 4.5$ GHz \cite{fangyu}.
Still the spectrum of RGWs stretches to such high frequencies
with enough power,
depending on both the energy scale involved in
the specific inflationary models and the reheating process.
For the designing parameters  for the proposed detector,
its sensitivity is short by about 6 orders of magnitude.
As for the prototype loop waveguide detector \cite{cruise}
operating at a high frequency $100$ MHz,
the sensitivity at present is short by much more.

\begin{figure}
\centerline{\includegraphics[width=10cm]{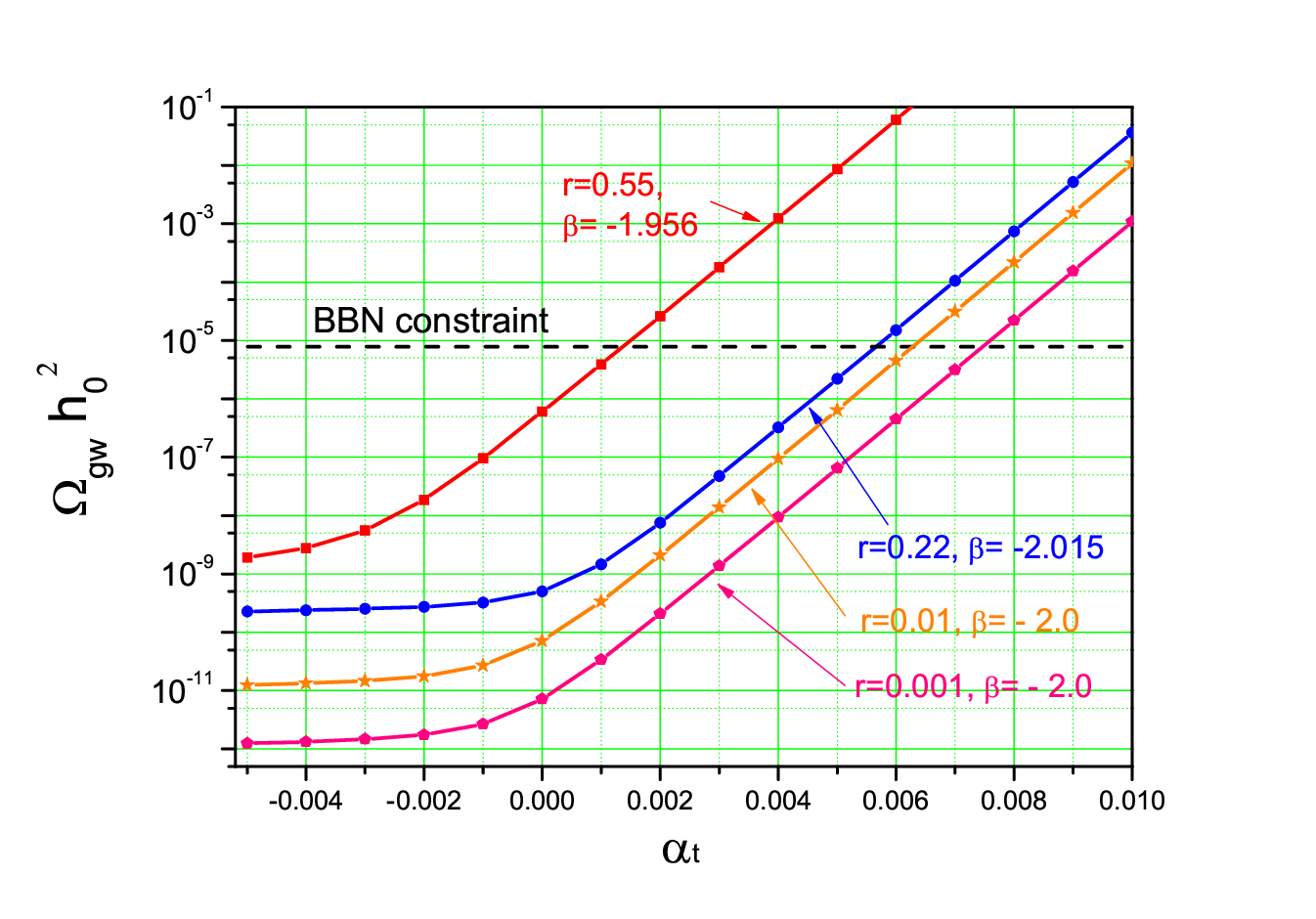}}
\caption{\label{BBN}
 The $\alpha_t$-dependence of energy density parameter
$\Omega_{gw}$ for various $r$ and $\beta$. }
\end{figure}

For lack of direct detection of RGWS,
cosmological considerations can be more effective
in providing the constraints upon the RGWs through its energy density.
In particular, the BBN process
occurring at a temperature $T\sim$ a few MeV during the early universe
is sensitive to the total cosmic energy density,
including that of RGWs.
An increase in the energy density of RGWs will enhance
the freezing temperature of neutrons and hence the light-element abundances.
Consequently, this will bring
observational constraints on the running index $\alpha_t$.
The energy density of RGWs should not be too large,
otherwise, it will significantly affect the outcome of the BBN process.
Measured abundances of light-element constrain
the number of additional relativistic species at BBN
to an equivalence of $\delta N_\nu=1.4$ neutrino degrees of freedom,
corresponding to $h_0^2\Omega_{gw}< 7.8\times 10^{-6}$
\cite{Cyburt,Maggiore}.
Besides,
if the initial perturbation amplitude of RGWs is non-adiabatic,
as is the case of RGWs being generated during inflation,
measurements of the CMB power spectrum also provide a constraint
$h_0^2  \Omega_{g}(\nu)  < 8.4\times 10^{-6}$
in the very low frequency range $\nu=(10^{-15},10^{-10})$ Hz \cite{smith},
comparable to that from BBN.
In constraining the energy density of RGWs,
sometimes $\Omega_{g}(\nu)$ and $ \Omega_{gw}$
were used interchangeably in literature.
But it is at most an approximation,
valid only under the condition that
the integration interval $d\nu/\nu=d \log \nu\sim 1$ and
$\Omega_{g}(\nu)$ is rather flat and smooth.
In this paper we distinguish $\Omega_{g}(\nu)$ and $ \Omega_{gw}$,
and give a more accurate treatment.
Using the normalization of Eq.(\ref{alpha})
and carrying out the integration in Eq. (\ref{gwe}),
we obtain the $\alpha_t-$dependence
of the the energy density parameter $\Omega_{gw}$
in Fig.\ref{BBN} for various values of parameters.
Adopting the BBN bound, the constraint on $\alpha_t$ is found to be
$\alpha_t\leq0.0014$ for $r=0.55$ and $\beta= -1.956$,
$\alpha_t\leq0.0056$ for $r=0.22$ and $\beta= -2.015$,
and $\alpha_t\leq0.0077$ for $r=0.001$ and $\beta= -2.0$, respectively.
One can infer that
the constraint  $\alpha_t <  0.008$
for any reasonable set of cosmological parameters.
It should be mentioned that,
although the BBN bound
yields a  constraint on RGWs,
it does not provides a direct detection of RGWs,
so one can only get upper bounds of $\alpha_t$.

This rather stringent constraint on $\alpha_t$
supports a nearly power-law spectrum of RGWs,
and is consistent with the expectation from
scalar inflationary models  \cite{Kosowsky}.
However, a comparison shows that,  the magnitude of $\alpha_t$
constrained by the BBN bound  is even smaller by one order
than the scalar running index $\alpha_s$ obtained by WMAP on large scales
\cite{Peiris,Spergel07,Komatsu,Dunkley}.
If both  RGWs and scalar perturbations are
generated by the same inflation,
one expects  $\alpha_s$ to be nearly as small as $\alpha_t$
for several kinds of smooth scalar potential \cite{Kosowsky}.
In light of the stringent constraint on $\alpha_t$,
it is hinted that $\alpha_s$ should also be rather small.
There has been some debate on the significantly non-vanishing
scalar running $\alpha_s$ \cite{Easther}.
Therefore, it is much desired that
constraints on the scalar running index $\alpha_s$ be drawn
from observations on smaller sales.

\begin{figure}
\centerline{\includegraphics[width=10cm]{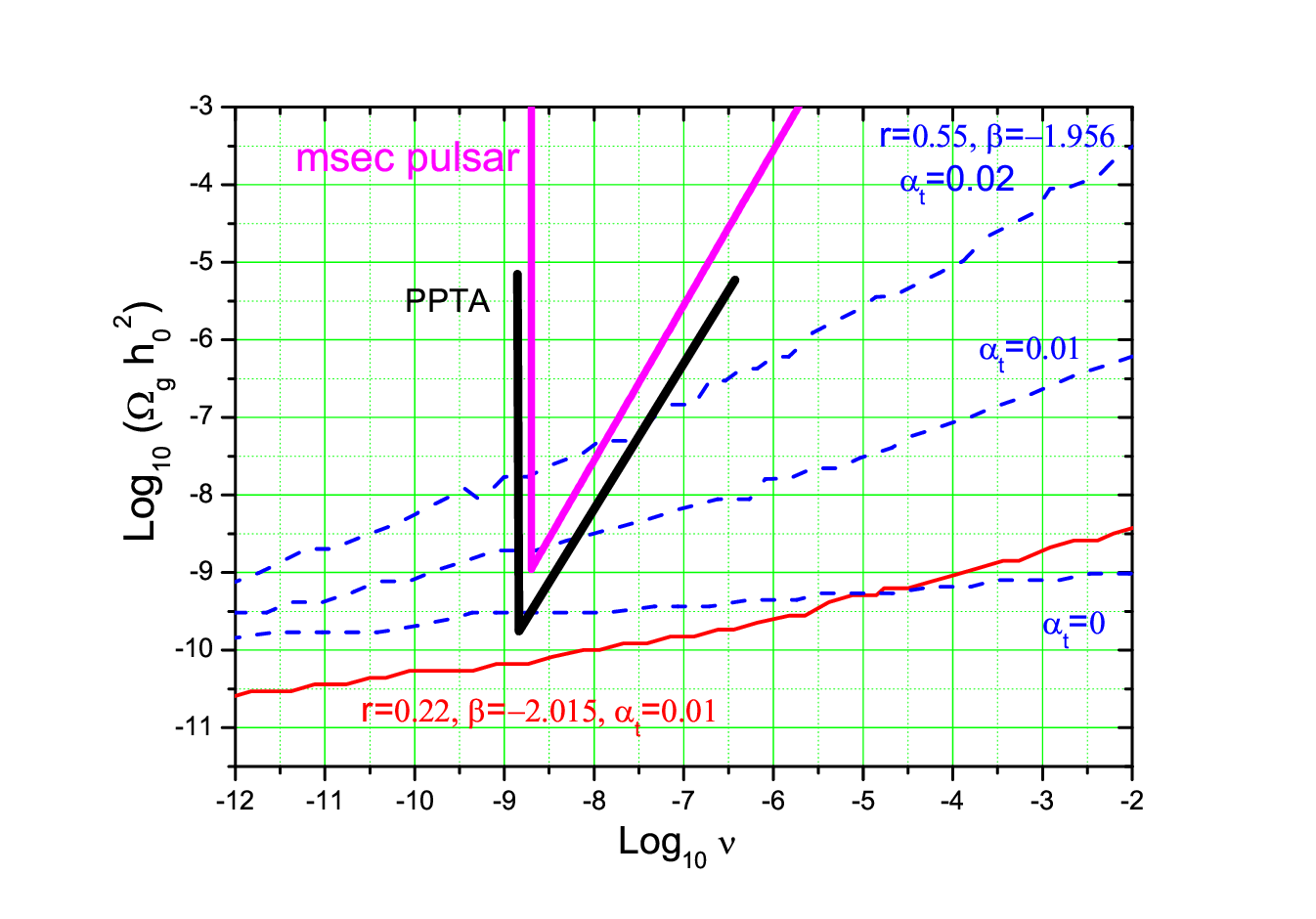}}
\caption{\label{pulsar}
Comparison with the constraint from data of millisecond pulsars
\cite{Lommen} and PPTA \cite{Manchester}. }
\end{figure}

Another kind of stringent constraints on RGWs
come from observations of millisecond pulsars,
which can serve as a gravitational wave detector
in the low frequency range.
For instance, by analyzing the uncertainty $\epsilon$
in the arrival timing of pulses for a duration $T$ of observation,
the pulsar will be sensitive to gravitational waves
$h(\nu,\tau_H)\sim \epsilon/T$ with $\nu\sim 1/T$.
For PSR B1855+09, a bound has been given for RGWs \cite{Maggiore,Kaspi}:
$\Omega_g(\nu_*)h_0^2 <4.8\times 10^{-9}(\nu/\nu_*)^2$ for $\nu>\nu_*$,
where $\nu_*=4.4\times 10^{-9}$ Hz.
Another treatment gives
$\Omega_g(\nu_*)h_0^2 <2\times 10^{-9}(\nu/\nu_*)^2$ for $\nu>\nu_*$,
where $\nu_*=1.9\times 10^{-9}$ Hz \cite{Lommen}.
Applying this bound to compare with
the calculated energy density spectrum $\Omega_g(\nu)$,
we obtain the constraint on RGWs, shown in  Fig.\ref{pulsar}.
For the parameters $r=0.55$ and $\beta= -1.956$,
the pulsar detector puts a constraint  $\alpha_t < 0.01$,
which is less stringent than the BBN constraint.
As an extension of this technique,
Parkes Pulsar Timing Array (PPTA) \cite{PPTA,Manchester}
consists of a sample of 20 millisecond pulsars
distributed over the entire sky,
and correlations in the time residuals of pulsars
help to disentangle RGWs signals.
After five years of observation,
it will improve the sensitivity by about one order
over that from a single pulsar
and will have a chance to detect RGWs of
$r=0.55$, $\beta=-1.956$, and $\alpha_t=0$.
But for  $r=0.22$ and  $\beta=-2.015$,
the detector will still not be able to detect RGWs
for a small value of $\alpha_t$.

In summary, by allowing for a tensorial running index $\alpha_t$,
the RGWs spectrum $h(\nu,\tau_H)$ will be significantly affected,
particularly in higher frequencies.
A positive $\alpha_t$ tends to
bend the high frequency portion of $h(\nu,\tau_H)$ to upward,
while a negative $\alpha_t$ will do the opposite.
This has brought about significant consequences to
various ongoing and planned detectors,
and led to reexaminations of the constraints on parameters of RGWs
that were previously estimated for the case $\alpha_t=0$.
For instance, a small variation of $\alpha_t$
from $-0.01$ to $0.01$ will increase the amplitude RGWs
by several orders of magnitude,
depending on frequencies.
It is interesting to note that
LISA by its design will be able to detect RGWs with $\alpha_t \ge 0$
for  the parameters $r = 0.22$, $\beta = -2.015$.
For an inflationary model with $r=0.55$ and  $\beta=-1.956$,
the observational constraint
is $\alpha_t<0.013$ from LIGO S5,
$\alpha_t<0.01$ from millisecond pulsar  PSR B1855+09.
The most stringent constraint coming from the BBN bound
is $\alpha_t<0.008$ as a rather conservative estimate
for any reasonable set of cosmological parameter ($\beta$, $r$).
The resulting tiny $\alpha_t$ prefers
the simple inflationary models with a nearly power-law spectrum of RGWs,
and would also hint a rather small scalar running index $\alpha_s$
within scalar inflationary models.
It is also found that
there is a degeneracy of $\alpha_t$ with  $\beta$ and $r$
in a narrow interval of detection frequency.
Detectors working in a broad frequency range,
such as LISA, may have a better chance in breaking the degeneracy.

~
~

ACKNOWLEDGMENT:
M. L. Tong's work has been partially supported by Graduate
Student Research Funding from USTC. Y. Zhang's work has
been supported by the CNSF No. 10773009, SRFDP, and CAS.

\end{document}